\documentclass[twocolumn,amsmath,amssymb,nofootinbib,tighten,floatfix,prb]{revtex4}

\usepackage{graphicx}
\usepackage{dcolumn} 
\usepackage{bm, bbm}      
\usepackage{curves}
\usepackage{epic}
\usepackage{wasysym}
\usepackage{epsf, times}
\usepackage{subfigure}
\usepackage{eufrak}
\usepackage{amsthm}

\def\openone{\leavevmode\hbox{\small1\kern-4.2pt\normalsize1}}

\newcommand{\bfig}{\begin{figure}}
\newcommand{\efig}{\end{figure}}
\newcommand{\beq}{\begin{equation}}
\newcommand{\eeq}{\end{equation}}
\newcommand{\bea}{\begin{eqnarray}}
\newcommand{\eea}{\end{eqnarray}}

\newcommand{\br}{{\bf r}}

\newcommand{\Q}{q}

\def\tit#1#2#3#4#5{{#1}{\bf #2}, #3 (#4)}

\def\prl{Phys.\ Rev.\ Lett.\ }

\def\prb{Phys.\ Rev.\ B\ }

\def\rpp{Rep.\ Progr.\ Phys.\ }

\begin{document}

\title{
Magnetic Monopoles in Spin Ice
      }

\author{C.~Castelnovo$^1$, R.~Moessner$^{1,2}$, and S.~L.~Sondhi$^3$}

\affiliation{$^1$Rudolf Peierls Centre for Theoretical Physics, 
Oxford University, Oxford OX1 3NP, UK}
\affiliation{$^2$Max-Planck-Institut f\"ur Physik komplexer Systeme, 
01187 Dresden, Germany }
\affiliation{$^3$Department of Physics, Princeton University, 
Princeton, NJ 08544}

\date{\today}

\begin{abstract}
Electrically charged particles, such as the electron, are
ubiquitous. By contrast, no elementary particles with a net magnetic
charge have ever been observed, despite intensive and prolonged
searches\cite{monosearch}. We pursue an alternative strategy, namely
that of realising them not as elementary but rather as {\em emergent}
particles, i.e., as manifestations of the correlations present in a
strongly interacting many-body system. The most prominent examples of
emergent quasiparticles are the ones with fractional electric charge
$e/3$ in quantum Hall physics\cite{laughlinnobellecture}.  Here we
show that magnetic monopoles {\em do} emerge in a class of exotic
magnets known collectively as spin
ice\cite{harbram97,ramirezspinice,bgr}: the dipole moment of the
underlying electronic degrees of freedom fractionalises into monopoles. 
This enables us
to account for a mysterious phase transition observed experimentally
in spin ice in a magnetic field\cite{kagice1,kagice2}, which is a liquid-gas
transition of the magnetic monopoles.  These monopoles can also be
detected by other means, e.g., in an experiment modelled
after the celebrated Stanford magnetic monopole
search\cite{stanfordmono}.
\end{abstract}

\maketitle
%
%

\begin{figure}
\includegraphics[width=0.99\columnwidth]{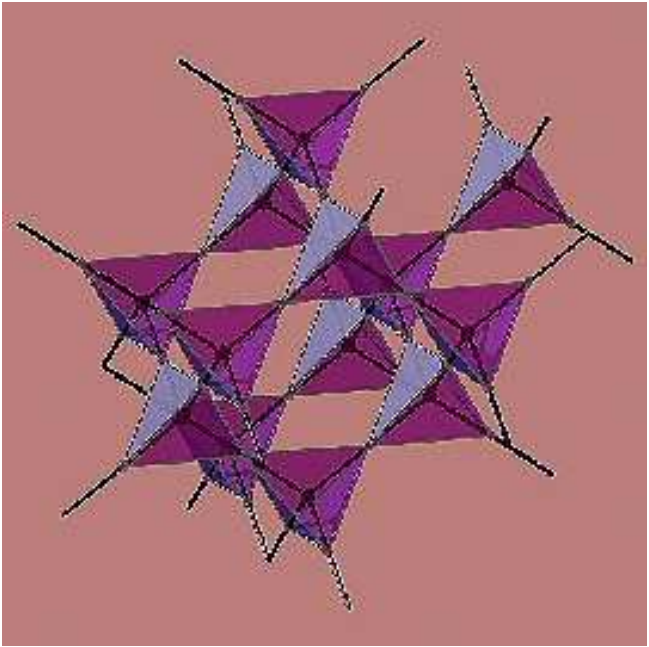}
\caption{
\label{fig:lattices}
The pyrochlore and diamond lattices. 
The magnetic moments 
in spin ice reside on the sites of the pyrochlore lattice, which 
consists of corner-sharing tetrahedra.  
These are at the same time the midpoints of the bonds of the
diamond lattice (black) formed by the centres of the tetrahedra. 
The ratio of the lattice constant of the diamond and pyrochlore lattices 
is $a_d/a=\sqrt{3/2}$. 
The Ising axes are the local $[111]$ directions, which
point along the respective diamond lattice bonds. 
}
\end{figure}

Spin-ice materials are characterised by the presence of magnetic moments 
$\vec{\mu}_i$ residing on the sites of a pyrochlore lattice (depicted in 
Fig.~\ref{fig:lattices}). 
These moments are constrained to point along their 
respective local Ising axes $\hat{e}_i$ (the diamond lattice bonds 
in Fig.~\ref{fig:lattices}), and they can be modelled as Ising spins 
$\vec{\mu}_i = \mu S_i$, where $S_i=\pm1$ and $\mu = \vert\vec{\mu}_i\vert$. 
For the spin ice compounds discussed here, Dy$_2$Ti$_2$O$_7$ and 
Ho$_2$Ti$_2$O$_7$, the magnitude $\mu$ of the magnetic moments equals 
approximately $10$ Bohr magnetons ($\mu = 10 \mu_B$). 
The thermodynamic properties of these compounds are known to be described 
with good accuracy by an energy term that accounts for 
the nearest neighbour exchange and the long ranged dipolar
interactions\cite{siddener,ging112,bgr}: 
\bea
H
&=&
\frac{J}{3} \sum_{\langle ij \rangle}
S_i S_j 
\nonumber \\
&+& 
D a^{3}_{\ } 
\sum_{(ij)}
\left[
  \frac{\hat{e}^{\ }_{i} \cdot \hat{e}^{\ }_{j}}
       {\vert \br^{\ }_{ij} \vert^{3}_{\ }}
  -
  \frac{3 \left(\hat{e}^{\ }_{i} \cdot \br^{\ }_{ij}\right)
          \left(\hat{e}^{\ }_{j} \cdot \br^{\ }_{ij}\right)}
       {\vert \br^{\ }_{ij} \vert^{5}_{\ }}
\right]
S_i S_j 
. 
\nonumber \\ 
\label{eq:dipE}
\label{eq: spin ice energy}
\eea
The distance between spins is $r_{ij}$, and $a \simeq 3.54$ \AA\ is the 
pyrochlore nearest-neighbour distance.
$D = \mu_0\mu^2 / (4 \pi a^3)=1.41$K is the coupling constant of the dipolar interaction. 

Spin ice was identified as a very unusual magnet when it was noted
that it does not order to the lowest temperatures, $T$, even though it
appeared to have 
{\em ferromagnetic} interactions\cite{harbram97}. Indeed, spin ice
was found to have a residual entropy at low $T$~\cite{ramirezspinice}, 
which is well-approximated by the famous Pauling entropy
for (water) ice, $S\approx S_{P}=(1/2) \log (3/2)$ per spin.
Pauling's entropy measures the huge ground state degeneracy arising from
the so-called ice rules. In the context of spin ice, its observation implies
a macroscopically degenerate ground state manifold obeying the ``ice rule''
that two spins point into each vertex of the diamond lattice, and two out. 

We contend that excitations above this ground state manifold, i.e.,\
defects that locally violate the ice rule, are magnetic monopoles
with the necessary long distance properties. From the perspective of
the seemingly local physics of the ice rule, the emergence of monopoles
would at first sight seem rather surprising.
To demystify it we will probe deeper into how the long
range magnetic interactions contained in Eq.~\ref{eq: spin ice energy}
give rise to the ice rule in the first place. We then incorporate insights
from recent progress in understanding the entropic physics of spin
ice, and the physics of fractionalisation in high 
dimensions\cite{hkms,hermele,henley,fulde,quantether}, 
of which our monopoles will prove to be a classical instance.

\begin{figure}
\includegraphics[width=0.49\columnwidth]{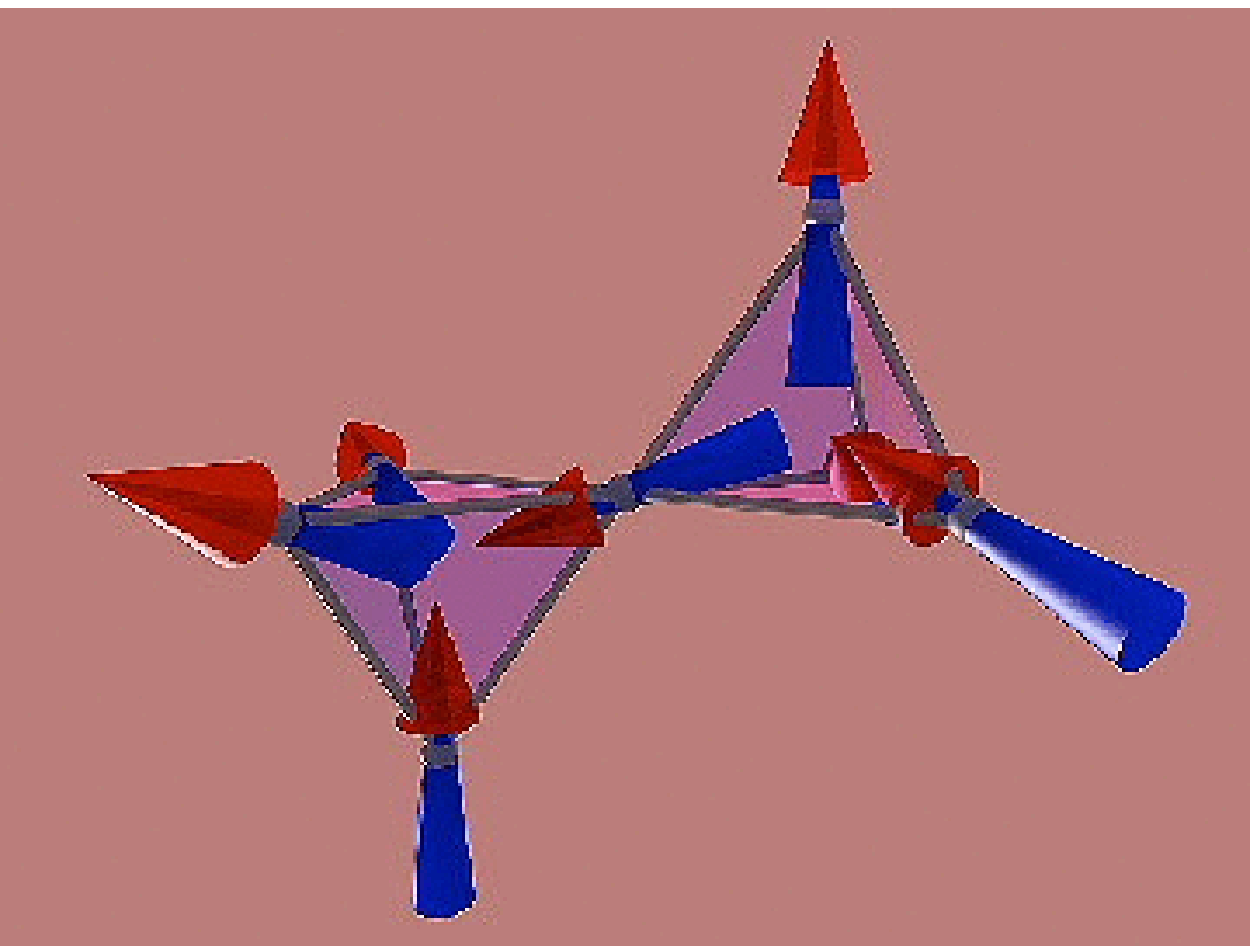}
\includegraphics[width=0.49\columnwidth]{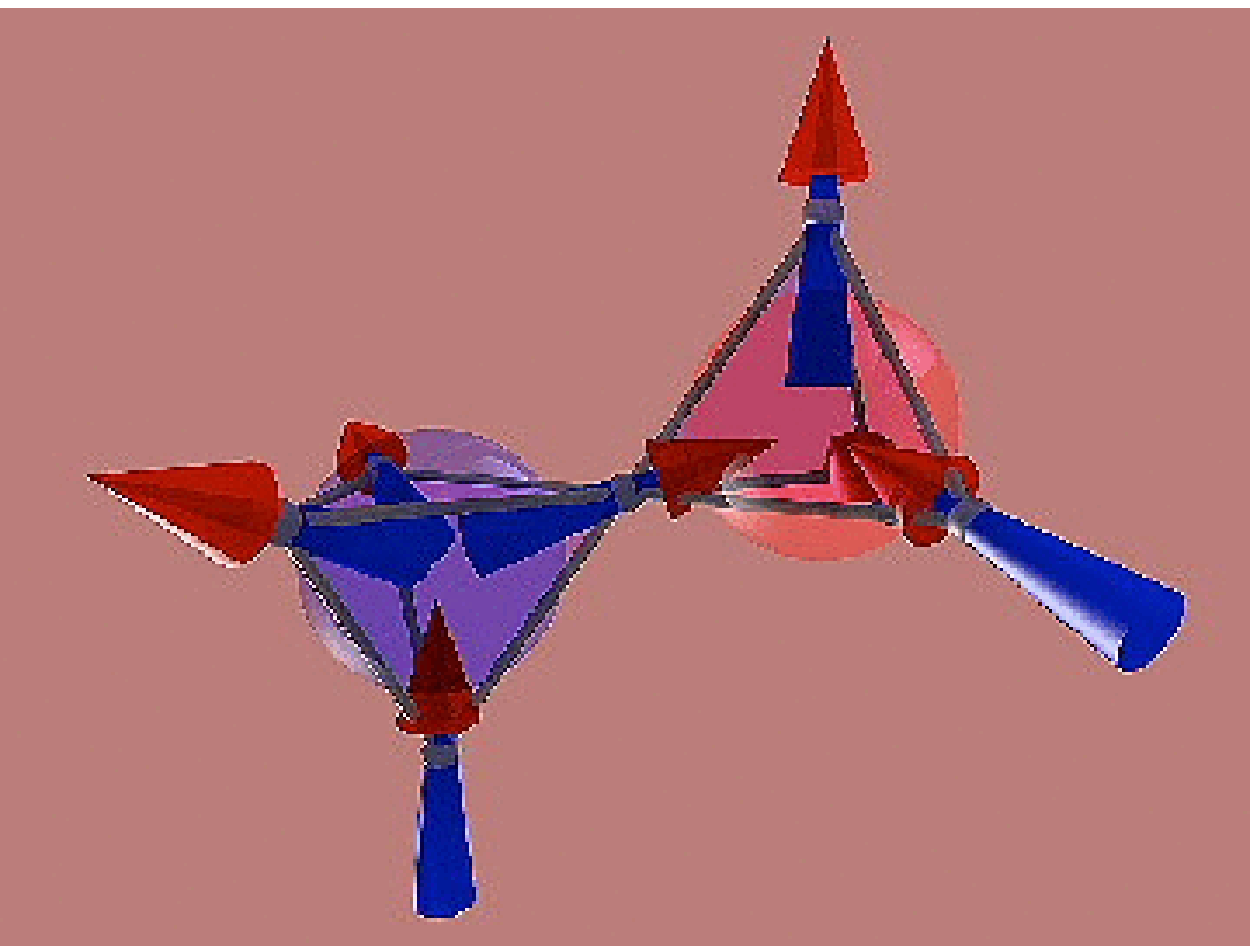}
\\ 
\includegraphics[width=0.49\columnwidth]{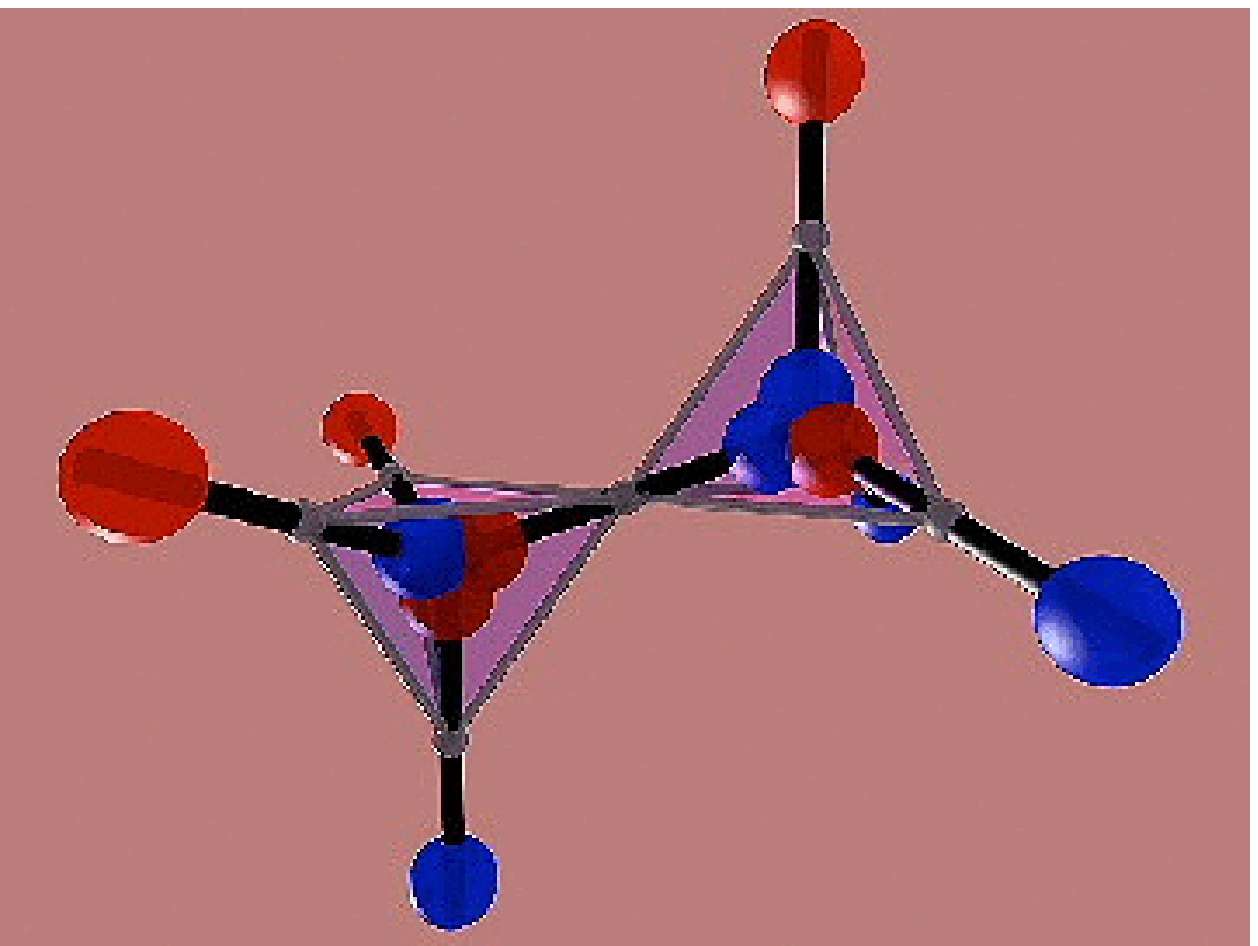}
\includegraphics[width=0.49\columnwidth]{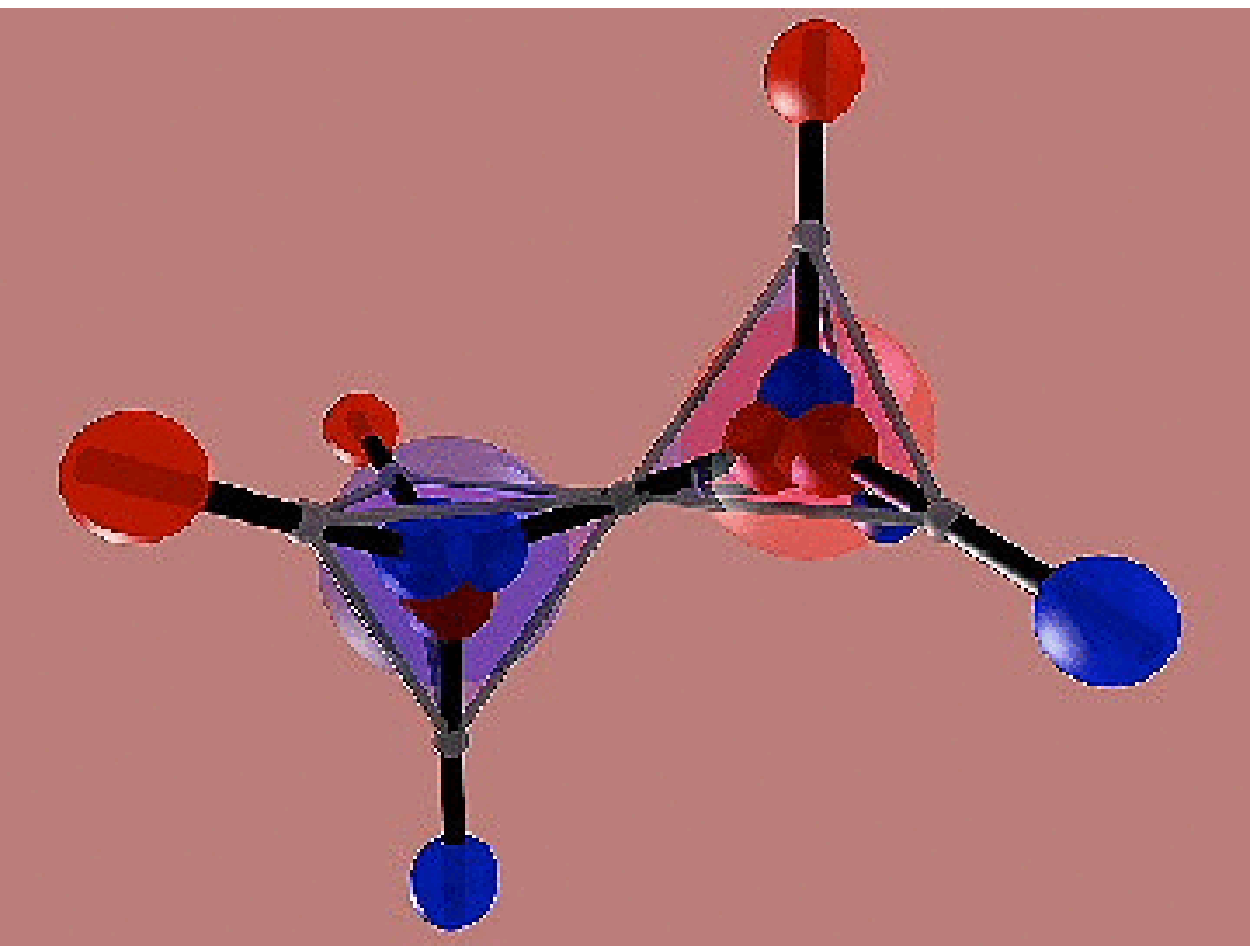}
\\ 
\includegraphics[width=0.99\columnwidth]{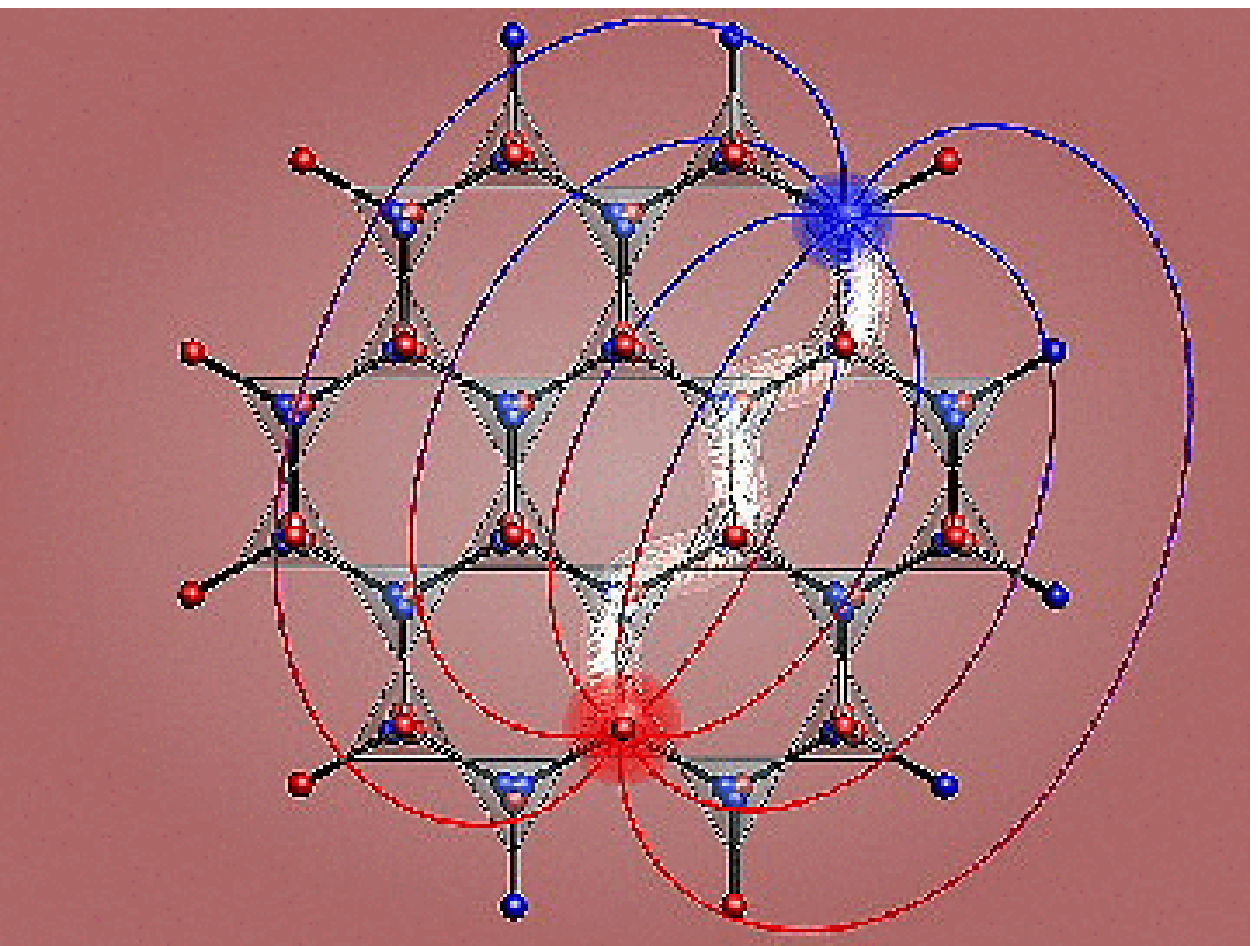}
\caption{
\label{fig: dipole to dumbbell} \label{fig: defects}\label{fig:pyro2kag}
Mapping from dipoles to dumbbells.  The dumbbell picture (c,d)
is obtained by replacing each spin in (a,b) by a pair of opposite
magnetic charges placed on the adjacent sites of the diamond lattice.
Left: two neighboring tetrahedra obeying the ice rule, with
two spins pointing in and two out, giving zero net charge on each site. 
Right: Inverting the shared spin generates a pair of magnetic monopoles 
(diamond sites with net magnetic charge). 
This configuration has a higher net magnetic moment and it is favoured by 
an applied magnetic field oriented upward in the figure 
(corresponding to a $[111]$ direction). 
Large panel: A pair or separated monopoles (large red and blue spheres).
A chain of inverted dipoles (`Dirac string')
between them is highlighted, and the magnetic field lines are sketched.
}
\end{figure}

Consider a modest deformation of Eq.~\ref{eq: spin ice energy},
loosely inspired by Nagle's old work\cite{nagleunit} 
on the `unit model' description of water ice: 
we replace the interaction energy of the magnetic 
dipoles living on pyrochlore sites by that of dumbbells 
consisting of equal and opposite magnetic charges that live at the ends of 
the diamond bonds (see Fig.~\ref{fig: dipole to dumbbell}). 
The two ways of assigning charges on each diamond bond reproduce the two 
orientations of the original dipole. 
Demanding that the dipole moment of the spin be
reproduced quantitatively fixes the value of the charge at $\pm\mu/a_d$, 
where the diamond lattice bond length $a_d = \sqrt{3/2}\,a$. 

The energy of a configuration of dipoles is computed as the 
pairwise interaction energy of magnetic charges, given by the magnetic 
Coulomb law 
\bea
V(r_{\alpha\beta}) 
&=& 
\left\{
  \begin{array}{ll}
    \frac{\mu_0}{4\pi} \frac{Q_\alpha Q_\beta}{r_{\alpha\beta}}
    & 
    \alpha \neq \beta 
    \\ 
    \frac{1}{2} v_o Q^2_\alpha 
    & 
    \alpha = \beta, 
  \end{array}
\right. 
\label{eq:monE}
\eea
where $Q_\alpha$ denotes the total magnetic charge at site $\alpha$ in
the diamond lattice, and $r_{\alpha\beta}$ is the distance between two 
sites. 
The finite ``self-energy'' $v_0 / 2$ is required to 
reproduce the net nearest neighbour interaction correctly. 
This expression -- derived in detail in the annexed supplementary notes -- 
is equivalent to 
the dipolar energy Eq.~\ref{eq:dipE} up to corrections which are small
everywhere, and vanish with distance at least as fast as $1/r^{5}$.

Consider first the ground states of the system. The total energy is
minimised if each diamond lattice site is net neutral, i.e., we must
orient the dumbbells so that $Q_\alpha = 0$ on each site. But this is
just the above-mentioned ice rule, as illustrated in Fig.~\ref{fig:
dipole to dumbbell}.  Thus, one of the most remarkable features of spin
ice pops right out of the dumbbell model: the measured low-$T$ entropy
agrees with the Pauling entropy (which follows from the short-distance
ice rules), even though the dipolar interactions are long-ranged.

\begin{figure}
\includegraphics[width=0.99\columnwidth]{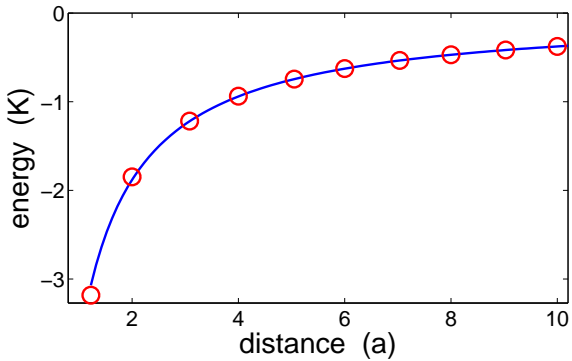} 
\caption{
\label{fig: Coulomb check} 
Monopole interaction.
Comparison of the magnetic Coulomb 
energy $-\mu_0\Q_m^2/(4\pi
r)$ (Eq.~\ref{eq:monE}, solid line) with 
a direct numerical evaluation of the
monopole interaction energy in 
dipolar spin ice (Eq.~\ref{eq:dipE}, open circles), for a given
spin-ice configuration (Fig.~\ref{fig: defects}), as a function of
monopole separation.
}
\end{figure}

We now turn to the excited states.  Naively, the most elementary
excitation involves inverting a single dipole~/~dumbbell to generate a
local net dipole moment $2\mu$.  However, this is misleading in a
crucial sense. The inverted dumbbell in fact corresponds to two
adjacent sites with net magnetic charge 
\beq
Q_\alpha =\pm \Q_m= \pm 2\mu/a_d 
, 
\eeq
a nearest neighbour monopole-antimonopole pair.  As shown
in Fig.~\ref{fig: defects}, the monopoles can be separated from one
another without further violations of local neutrality by flipping a
chain of adjacent dumbbells. A pair of monopoles separated by a
distance $r$ experiences a Coulombic interaction, $-\mu_0\Q_m^2/(4\pi
r)$, mediated by monopolar magnetic fields, see 
Fig.~\ref{fig: Coulomb check}. 

This interaction is genuinely magnetic, hence the presence of 
the vacuum permeability $\mu_0$,
and not $1/\epsilon_0$. It takes only a finite energy to
separate the monopoles to infinity (i.e., {\em they are deconfined}),
and thus they are the true elementary excitations of the system: the
local dipolar excitation fractionalizes!

The reader may now feel that we have slipped the baby in
with the bathwater -- that in taking the pictures from the dumbbell
representation seriously we are somehow introducing monopoles where
there were none to begin with. In general,  
it is of course well known that a string of dipoles
arranged head to tail realises a monopole-antimonopole pair at its
ends\cite{jacksonmono}.
However, in order to obtain {\em deconfined} monopoles, it is
essential that the cost of creating such a string of dipoles remain
bounded as its length grows, 
i.e., that the relevant string tension vanish.
This is evidently not true in a vacuum (such as that of the universe)
where the growth of
the string can only come at the cost of creating additional
dipoles. Magnetic materials, which come equipped with
vacua (ground states) filled with magnetic dipoles, are more promising. 
However, even here a dipole string is not always a natural 
excitation, and when it is -- e.g., in an ordered ferromagnet -- 
a string of inverted dipoles is accompanied by costly domain walls along 
its length (except, as usual, for one-dimensional systems\cite{saitoh}), 
causing the incipient monopoles to remain confined. 

The magic of spin ice arises from its exotic ground states. The ice rule can
be viewed as requiring that two dipole strings enter and exit each site 
of the diamond lattice. 
In a typical spin ice ground state, there is a ``soup'' of such strings: 
many dipole strings -- of arbitrary size and shape -- can be identified 
which connect a given pair of sites. 
Inverting the dipoles along any one such string creates a 
monopole-antimonopole pair on the sites at its ends. 
The associated energy cost does not diverge with the length of the string,
unlike the case of an ordered ferromagnet, as no domain walls are created 
along the string, and the monopoles are thus deconfined. 

Notice that we did not make reference to
the famous Dirac condition\cite{Dirac1931} that the fundamental
electric charge $e$ and any magnetic charge $\Q$ must exhibit the
relationship $e \Q = n h / \mu_0$
whence any monopoles in our universe must be quantised in units of $
q_D = h / \mu_0 e$.
This follows from the monopole being attached to a Dirac
string which has to be un{\em observable}. \cite{jacksonmono} 
By contrast, the string
soup characteristic of spin ice at low temperature makes the strings
energetically un{\em important}, although they are observable and thus
not quantised.

Indeed, the monopoles in spin ice have a magnitude 
\bea
\Q_m 
&=& 
\frac{2 \mu}{a_d} 
\nonumber \\
&=& 
\frac{2 \mu}{\mu_b} \frac{\alpha\lambda_C}{2\pi a_d} q_D 
\nonumber \\
&\approx& 
\frac{q_D}{8000} 
, 
\eea
where $\lambda_C$ is the Compton wavelength for an electron, and $\alpha$ 
is the fine-structure constant. Amusingly, the charge of a monopole in
spin ice can even be tuned {\em continuously} by applying pressure, as
this changes the value of $\mu/a_d$.

The monopoles are sources and sinks of the magnetic field $\bf H$, 
as is appropriate to the condensed matter setting.
More precisely, like in other instances of fractionalisation\cite{rajafrac}, 
one can define a `smeared' magnetic charge 
%
%
\beq
\rho_m({\bf R}) 
= 
\int d^3{\bf r^\prime} \: 
  e^{-|{\bf r^\prime-R}|^2 / \xi^2} \: 
    \nabla\cdot{\bf H} 
, 
\eeq
where $\nabla\cdot{\bf H}$ is the divergence of the magnetic field. 
For a monopole at the origin, separated by
$L\gg\xi\gg a$ from any other monopoles, this gives $\rho_m(0)=\pm q_m$. 
The form of the magnetic induction $\bf B$ is also monopolar, but with the
important difference that a compensating flux travels along the (unquantised)
``Dirac string'' of flipped dipoles created along with the monopole
(see Fig.~\ref{fig:pyro2kag}). 

Our magnetic monopoles would in principle show up in one of the
best-known monopole searches, the Stanford experiment to detect
fundamental magnetic monopoles from cosmic radiation.  This experiment
is based on the fact that a long-lived current is induced in a
superconducting ring when a monopole passes through
it\cite{stanfordmono}. One can easily check that the presence of the
Dirac string of flipped dipoles is immaterial to the establishment of a 
current.

The above observations are the central qualitative results of our
work: ice-rule-violating defects are deconfined monopoles of $\bf H$,
they exhibit a genuine magnetic Coulomb interaction,
Eq.~\ref{eq:monE}, and they produce Faraday electromotive
forces in the same way as elementary monopoles would.

We reemphasize that the ice rule alone does not permit a consistent
treatment of the excited states of the physical problem: crucially, the 
energetic interaction between our defects is absent altogether. 
Also, in previous discussions of the purely ice-rule problem and related 
short ranged problems\cite{hkms,hermele,henley} it has been noted that 
the defects do acquire a purely {\it entropic} Coulombic (i.e., $1/r$) 
interaction that has a strength that vanishes proportionally to $T$ at low 
temperatures.
This interaction will be present in addition to the magnetic Coulomb
interaction discussed in this paper, and is clearly much smaller as
$T \rightarrow 0$. Also, it will {\it not} be accompanied by a magnetic
field, it will not renormalise the monopole charge, and it will not be felt 
by a stationary magnetic test particle 
embedded in the lattice but not attached to a lattice site. 

The most satisfactory way to demonstrate the presence of a monopole
would be to measure the force on magnetic test particles, say by a
Rutherford scattering experiment or by clever nanotechnological means.
Unfortunately, for lack of elementary magnetic monopoles, one would
have to choose dipoles as test particles, which significantly weakens
such signatures. 

An alternative strategy is to look for consequences of the presence of
magnetic monopoles in the {\em collective} behaviour of spin ice. This
is most elegantly achieved by applying a magnetic field in the $[111]$
crystallographic direction. Such field acts as a (staggered) chemical
potential, see Fig.~\ref{fig: dipole to dumbbell}, 
favouring the creation of monopoles of a given sign on
either sublattice of the diamond lattice.  

We thus have a tuneable lattice gas of magnetic monopoles 
on the diamond lattice.
The basic structure of the phase diagram as a
function of magnetic field and temperature can be inferred from a
series of papers by M.~E.~Fisher and collaborators\cite{mefcoulomb} in
the context of ionic lattice gases and Coulombic criticality. At high
$T$, there is no phase transition but a continuous crossover between
the high- and low-density phases as the chemical potential is varied. 
At low $T$, a first-order 
phase transition separates the two regimes. This transition 
terminates in a critical point at $(h_c,T_c)$, 
not unlike the liquid-gas transition of
water. 
This serves as a useful diagnostic as the liquid-gas transition
is absent for a nearest-neighbour spin-ice model, 
in which defects interact only
entropically. In that case, it is known that there cannot be a
first-order transition in the limit of low $T$~\cite{heillieb}. 

\begin{figure*}[!ht]
\includegraphics[width=1.7\columnwidth]{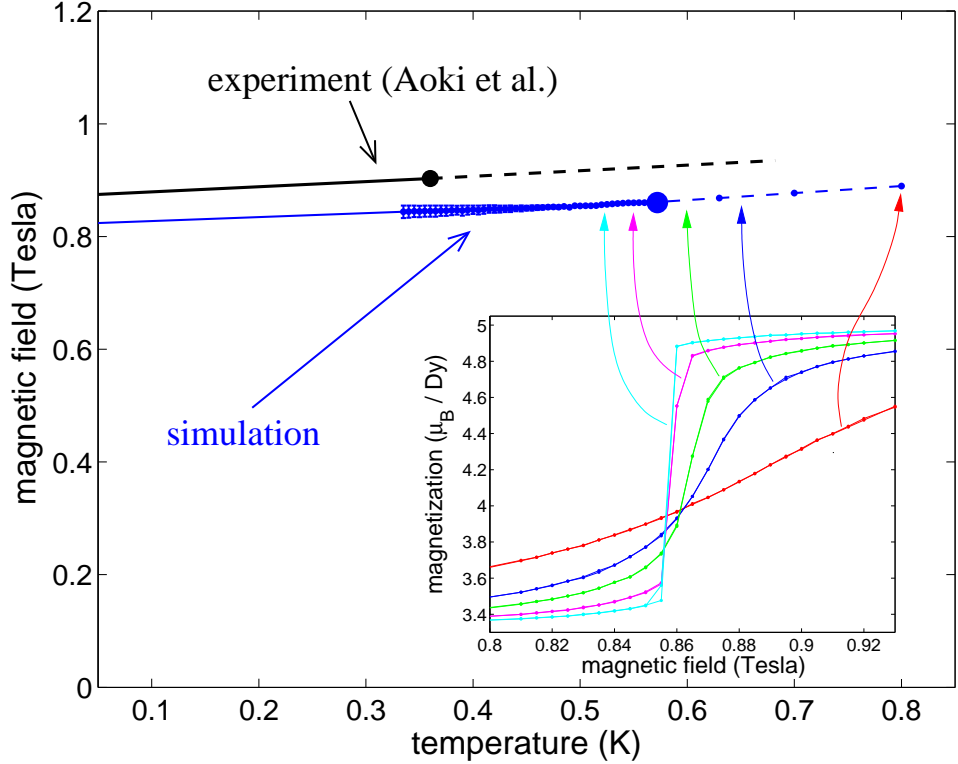}
\caption{\label{fig:liquidgasdiagram} 
Phase diagram of spin ice in a $[111]$ field.
The location of the monopole liquid-gas
transition from numerics (blue line) compared
to experiment (black line). The solid line is a
first-order transition terminated by a critical endpoint (fat circle).
The dashed lines are crossovers. 
Inset: magnetisation curves showing the onset
of first-order behaviour as the temperature is lowered.
Our simulations cover the range $0.335\,\textrm{K} <T< 0.8\,\textrm{K}$
for 1024 spins.
At the lowest temperatures, the parallel tempering code
we use in our  simulations of the Ewald-summed dipolar interaction
no longer completely suppresses
the hysteresis, and we have extended the first-order transition line 
using Clausius-Clapeyron.
}
\end{figure*}
To confirm this scenario, we have demonstrated by Monte
Carlo simulations that the actual phase diagram of  dipolar spin
ice model has precisely this structure. In
order to rule
out the appearance of the liquid-gas transition due to effects
introduced by the approximations leading to Eq.~\ref{eq:monE},
we simulated directly the original dipolar spin ice model, 
Eq.~\ref{eq: spin ice energy}.
The resulting phase diagram is depicted in Fig.~\ref{fig:liquidgasdiagram}. 
The critical endpoint is located
around $(T_c,h_c)=(0.57\pm0.06$K$,0.86\pm0.03$T$)$. The error bars are
mainly due to finite-size effects, as the intensive nature of the
simulations of long-range dipolar interactions prohibits simulating
very large systems.

This scenario is indeed observed experimentally in spin-ice 
materials\onlinecite{kagice1,kagice2}, 
and our results provide a natural explanation. 
Spin ice in a $[111]$ magnetic field is a very interesting problem 
which has already attracted considerable attention. The low-density phase of
monopoles is known as kagome ice, a quasi two-dimensional phase with algebraic 
correlations and a reduced residual entropy\cite{kagice1,kagice2,kagice3}. 
The high-density phase is an ordered state with maximal polarisation along the 
field direction. 
Experimental results on the liquid-gas transition and its endpoint
are also displayed in Fig.~\ref{fig:liquidgasdiagram} for comparison. Our
numerical results are in good qualitative agreement with both experiment
and the analytic calculations of Ref.~\onlinecite{mefcoulomb}. Our value
of the critical field agrees with Ref.~\onlinecite{kagice1} within
a few percent, less than the uncertainty due to 
demagnetisation effects\cite{kagice1,kagice2}. However, the experimental
value of $T_c$ is about a third lower than the numerical one, most likely 
due to farther-neighbour (exchange) interaction terms, which despite
being small can shift the location of a transition
temperature considerably\cite{ging112}. 

The presence of a liquid-gas transition was noted to be very
remarkable as there are few, if any, other experimentally known
instances in localised spin systems\cite{kagice1}. 
No mechanism was known to account for this phenomenon, and our theory of 
magnetic monopoles fills this gap. 

The existence of magnetic monopoles in a condensed matter system is, given 
their intellectual pedigree, exciting in itself. 
(The reader familiar with the beautiful work on the anomalous Hall effect 
will notice that the monopoles appearing there are not excitations and 
do not involve the physical magnetic field\cite{fn-fang}.) 
This is further heightened 
by their being a rare instance of high-dimensional fractionalisation, a 
phenomenon of great interest in fields as diverse as correlated electrons and 
topological quantum computing\cite{Kitaev}. 
We hope our analysis will encourage our experimentalist colleagues to try 
their hand at directly detecting these monopole. 
There are surely many avenues to explore in search of useful signatures: 
scattering, transport and noise measurements, flux detection, to mention 
just a few. 
%
%

\section*{
Acknowledgments
         }
The authors would like to thank S.~Bramwell, J.~Chalker, C.~Chamon and 
S.~Kivelson (especially for pointing out Ref.~\onlinecite{stanfordmono}) 
for discussions.  This work is supported in 
part by EPSRC Grant No.~GR/R83712/01 (CC), 
and NSF Grant No.~DMR 0213706 (SLS). 
The authors are deeply indebted to A.~Canossa for support with the graphics. 
%
%

\appendix 

\section{\label{sec: dumbell picture}
The dumbell picture
           }
This material presents a detailed derivation of the dumbell Hamiltonian used 
extensively in our paper. We start from the generally accepted Hamiltonian 
which contains a sum of nearest-neighbour exchange and long range dipolar 
interactions, 
\bea
H 
&=&
\frac{J}{3} \sum_{\langle ij \rangle}
S_i S_j 
\nonumber \\ 
&+& 
D a^{3}_{\ } 
\sum_{(ij)}
\left[
  \frac{\hat{e}^{\ }_{i} \cdot \hat{e}^{\ }_{j}}
       {\vert {\bf r}^{\ }_{ij} \vert^{3}_{\ }}
  -
  \frac{3 \left(\hat{e}^{\ }_{i} \cdot {\bf r}^{\ }_{ij}\right)
          \left(\hat{e}^{\ }_{j} \cdot {\bf r}^{\ }_{ij}\right)}
       {\vert {\bf r}^{\ }_{ij} \vert^{5}_{\ }}
\right]
S_i S_j 
. 
\nonumber \\
\label{SIeq: spin ice energy}
\eea
The magnetic moment of a spin is denoted by $\mu$, which equals 
approximately $10$ Bohr magnetons ($\mu=10 \mu_B$) for the spin ice compounds 
discussed here (namely, Dy$_2$Ti$_2$O$_7$ and Ho$_2$Ti$_2$O$_7$). 
The distance between spins is $r_{ij}$, and $a \simeq 3.54$ \AA\ is the 
pyrochlore nearest-neighbour distance.
$D = \mu_0\mu^2 / (4 \pi a^3)=1.41$K is the coupling constant of the dipolar 
interaction. 

\begin{figure}[!hb]
\includegraphics[width=0.49\columnwidth]{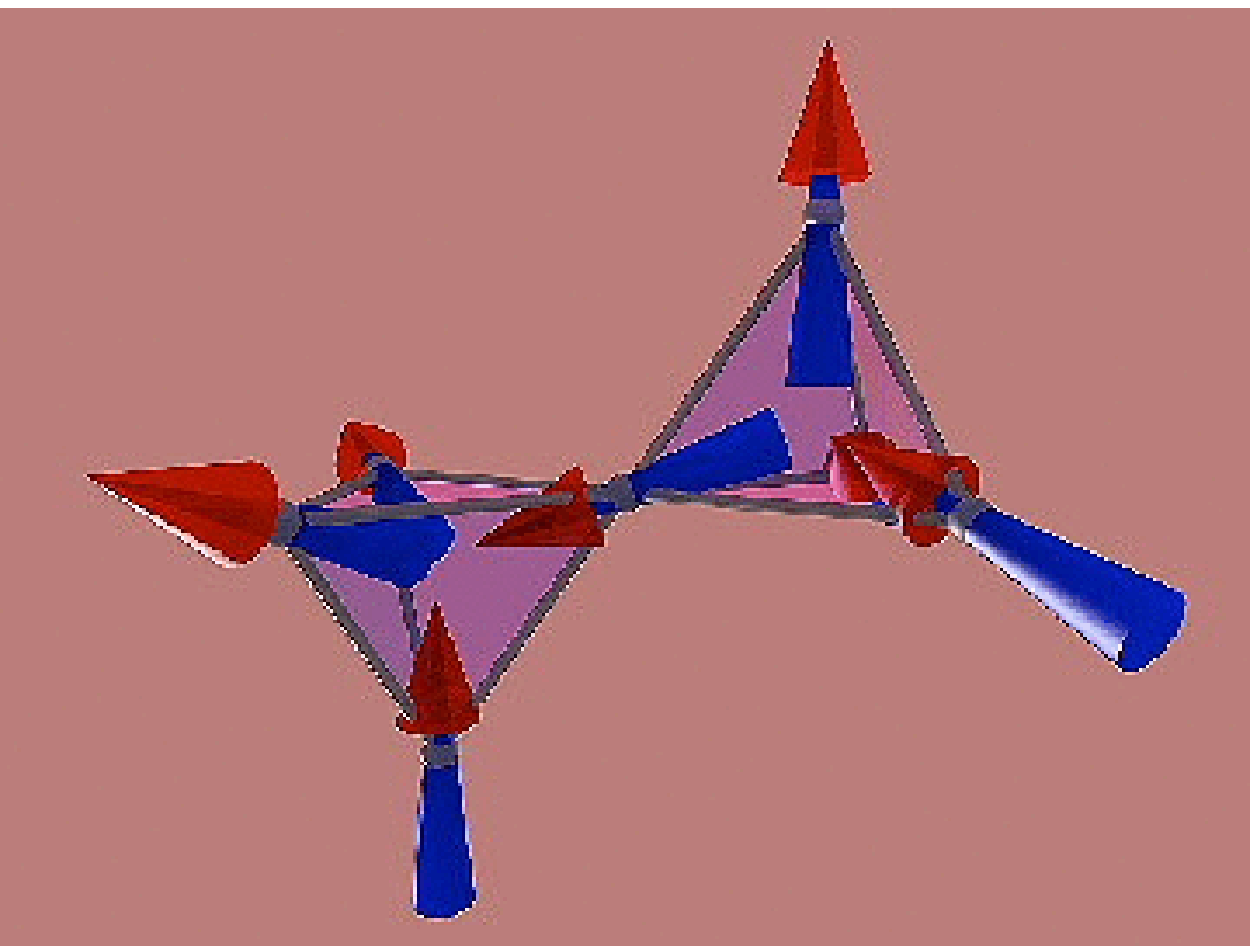}
\includegraphics[width=0.49\columnwidth]{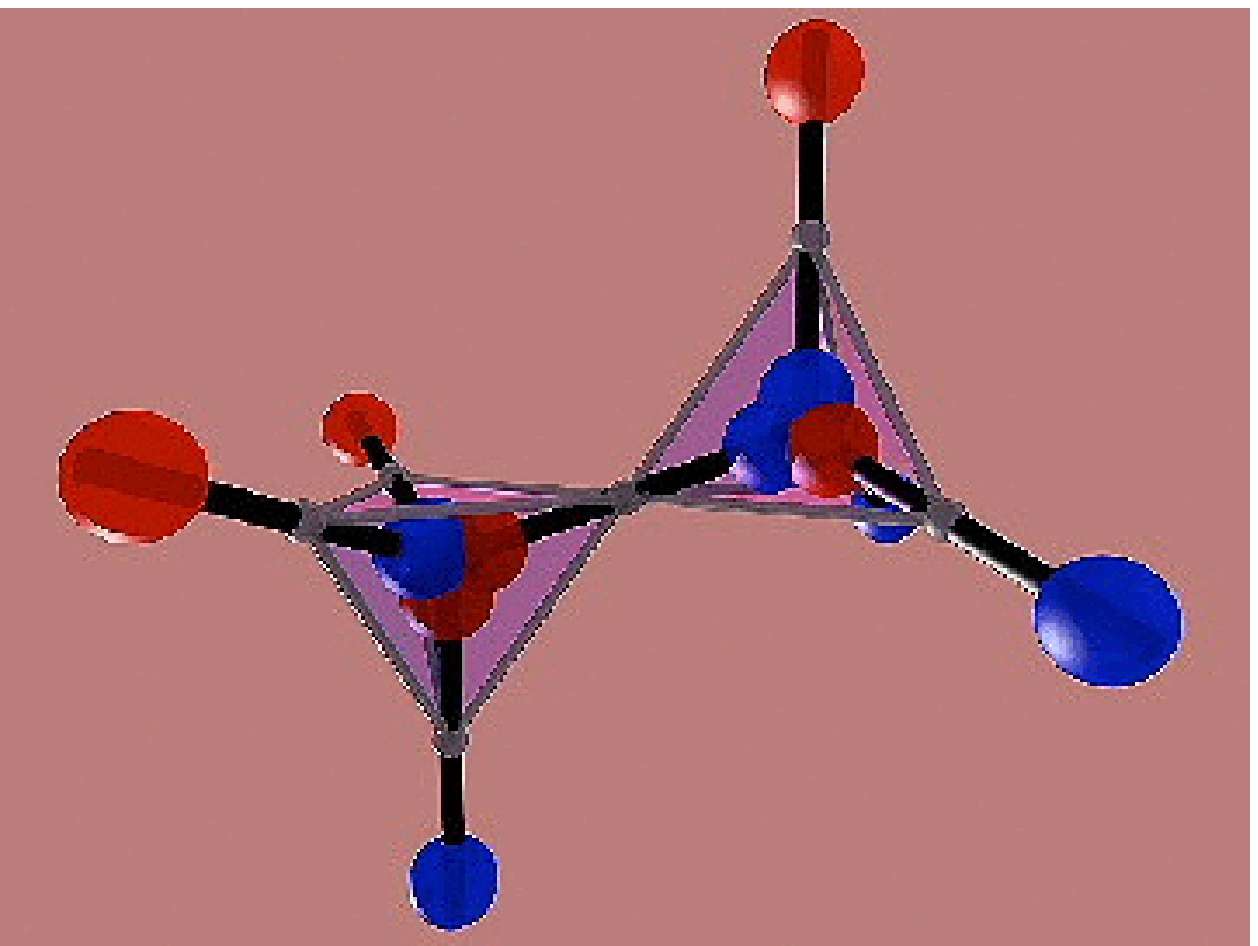}
\caption{
\label{SIfig: dipole to dumbell} 
Mapping from dipoles to dumbells. 
Left: two neighboring tetrahedra obeying the ice rule, with
two spins pointing in and two out, giving zero net charge on each site. 
Right: The corresponding dumbell picture obtained by replacing each spin by 
a pair of opposite magnetic charges placed on the adjacent sites of the 
diamond lattice. 
}
\end{figure}

A dipole can be thought of as a pair of equal and opposite charges of strength 
$\pm q$, separated by a distance $\tilde{a}$, such that $\mu = q \tilde{a}$. 
The dipolar part of the Hamiltonian is reproduced exactly in the limit 
$\tilde{a} \to 0$. 
Here we choose $\tilde{a}$ to equal the diamond lattice constant 
$a_d = \sqrt{3/2} \: a$, therefore fixing $q = \mu / a_d$. 
The two ways of assigning charges reproduce the two orientations of the
original dipole, as illustrated in Fig.~\ref{SIfig: dipole to dumbell}. 

\begin{figure}
\includegraphics[width=0.49\columnwidth]{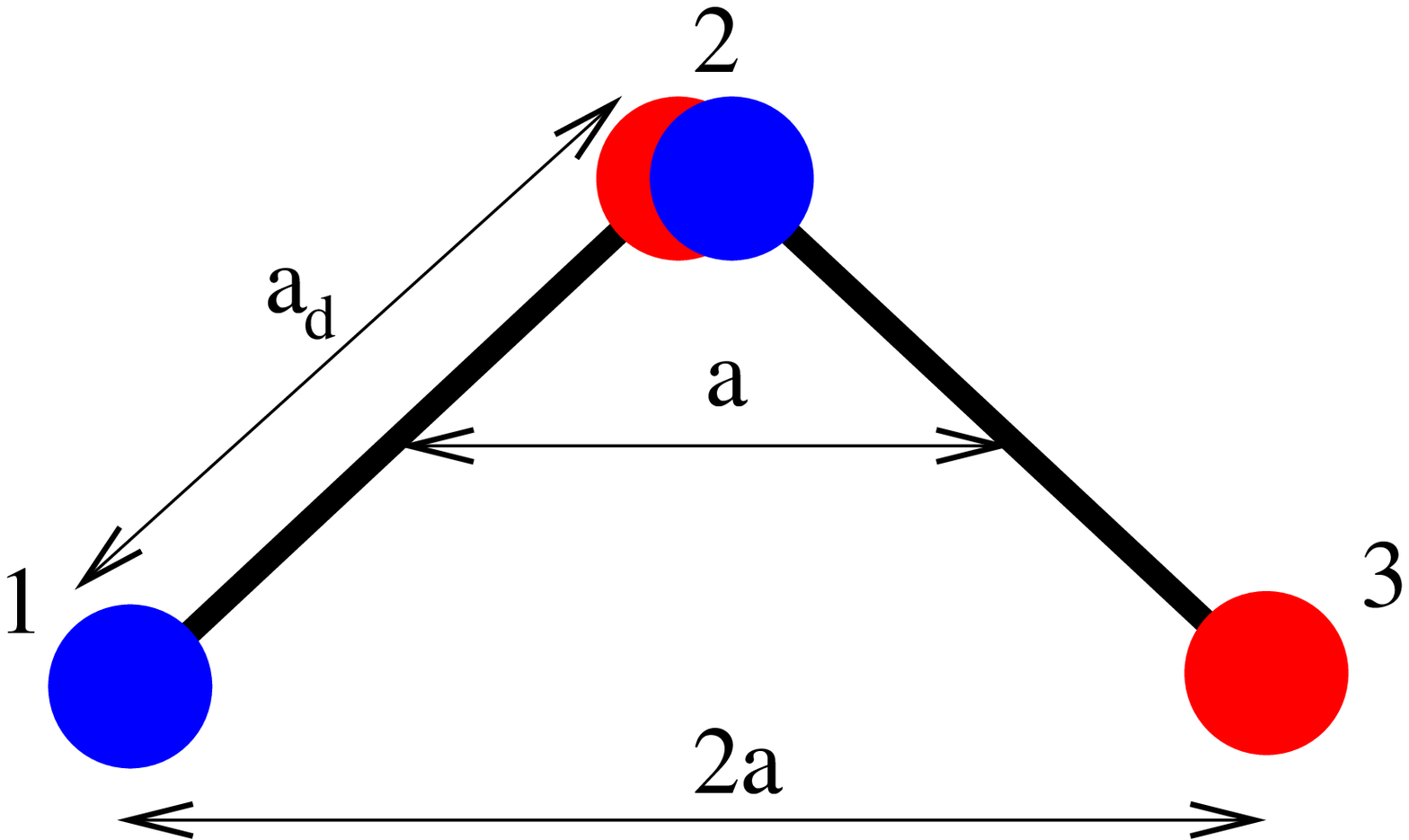}
\includegraphics[width=0.49\columnwidth]{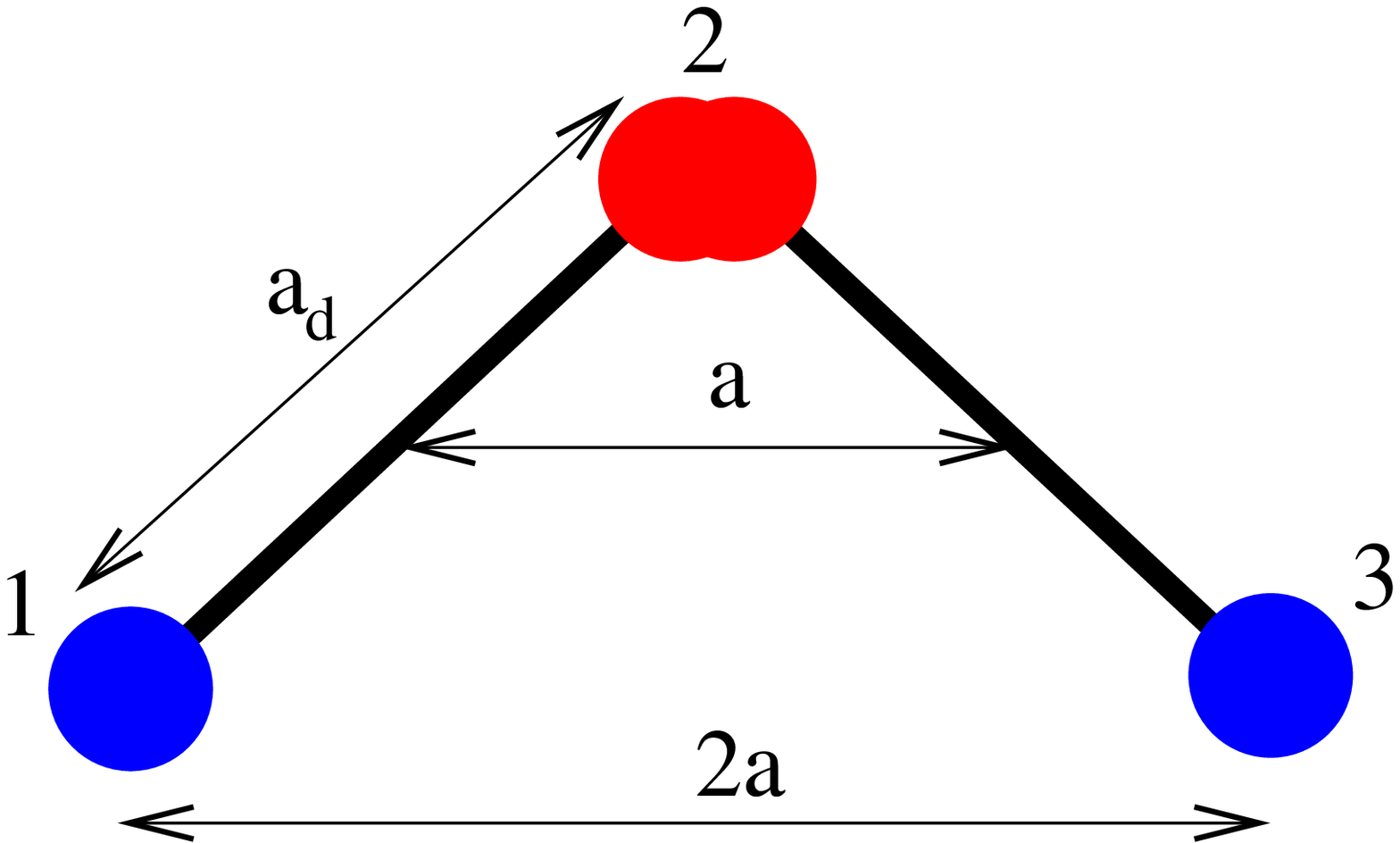}
\caption{
\label{SIfig: neighbouring dipoles} 
Schematic examples of two neighbouring dipoles on the pyrochlore lattice. 
}
\end{figure}

Given a configuration of $N$ dipoles, let us label 
$\{ q_i, \; i=1,\,\ldots,\,2N \}$ 
the $2N$ charges in the corresponding dumbell configuration. 
The magnetic Coulomb interaction between the charges is given by 
\bea
\mathcal{V}(r_{ij}) 
&=& 
\left\{
  \begin{array}{ll}
    \frac{\mu_0}{4\pi} \frac{q_i q_j}{r_{ij}}
    & 
    r_{ij} \neq 0 
    \\ 
    v_0 q_i q_j
    & 
    r_{ij} = 0 , 
  \end{array}
\right. 
\label{SIeq:chargeE}
\eea
which reproduces the interaction between two dipoles exactly in the limit of 
large separation between them (modulo an innocuous ```self-energy'' term 
discussed below). 
We can also tune the onsite contribution $v_0$ so as to match the interaction 
energy $\pm J_{\textrm{eff}} = \pm (J + 5D)/3$ between two neighbouring 
dipoles on the lattice (see for example 
Fig.~\ref{SIfig: neighbouring dipoles}). 
{}For two neighbouring spins both pointing into the shared tetrahedron 
(right panel of Fig.~\ref{SIfig: neighbouring dipoles}) one has 
\bea
J_{\textrm{eff}}
&=& 
v(0) - 2v(r_{12}) - 2v(r_{23}) + v(r_{13}) 
\nonumber \\ 
&=& 
v(0) - v(r_{12}) - v(r_{23}) + 2E_{\textrm{self}} + v(r_{13}) 
, 
\nonumber \\
\eea
where 
$
v(r) 
= 
\vert \mathcal{V}(r) \vert
$, 
and $E_{\textrm{self}} = -v(r_{12}) = -v(r_{23})$ is the self-energy of a 
dipole in the dumbell picture. 
For two neighbouring spins pointing into and out of the shared tetrahedron 
(as in the left panel of Fig.~\ref{SIfig: neighbouring dipoles}), 
we obtain 
\bea
- J_{\textrm{eff}}
&=& 
- v(0) + 2v(r_{12}) + 2v(r_{23}) - v(r_{13}) 
\nonumber \\ 
&=& 
- v(0) + v(r_{12}) + v(r_{23}) + 2E_{\textrm{self}} - v(r_{13}) 
. 
\nonumber \\
\eea
Taking the difference between the two equations above, and using the fact 
that $r_{12} = r_{23}$ and therefore $v(r_{12}) = v(r_{23})$, 
one gets  
\beq 
v(0) = J_{\textrm{eff}} + 2v(r_{12}) - v(r_{13}) 
\eeq
where 
\bea
r_{12} &=& a_d 
\nonumber \\ 
r_{13} &=& 2a
\nonumber \\ 
v(r) 
&=& 
\left\{
\begin{array}{ll}
  \frac{\mu_0}{4\pi} \frac{(\mu / a_d)^2}{r} & \qquad r \neq 0 
  \\ 
  v_0 \left( \frac{\mu}{a_d} \right)^2 & \qquad r = 0 
\end{array}
\right. 
\nonumber 
\eea
and in the end  
\beq 
v(0) 
= 
\frac{J+5D}{3}  
+  
\frac{\mu_0 }{4\pi} 
  \left(\frac{\mu }{a_d}\right)^2  
    \left[ \frac{2}{a_d} - \frac{1}{2a} \right] 
\eeq

Using the fact that $D = \mu_0 \mu^2 / 4 \pi a^3$ and that 
$a = a_d \sqrt{2/3}$, the expression above reduces to 
\bea
v_0 \: \left( \frac{\mu }{a_d} \right)^2 
= 
\frac{J}{3} + \frac{4}{3} \left[ 1 + \sqrt{\frac{2}{3}} \right] D . 
\label{SIeq: v_0}
\eea

Now, the energy of a charge configuration can be straightforwardly 
resummed and written in terms of the total charges $Q_\alpha$ on each site 
$\alpha$ of the diamond lattice, i.e., 
$Q_\alpha = q_{i_1} + q_{i_2} + q_{i_3} + q_{i_4}$ for the quartet of charges 
with $r_{i_1} = r_{i_2} = r_{i_3} = r_{i_4} \equiv r_\alpha$. 
The energy can thus be computed from a 
pairwise Coulomb interaction of the magnetic charges $Q_\alpha$: 
\bea
V(r_{\alpha\beta}) 
&=& 
\left\{
  \begin{array}{ll}
    \frac{\mu_0}{4\pi} \frac{Q_\alpha Q_\beta}{r_{\alpha\beta}}
    & 
    \alpha \neq \beta 
    \\ 
    \frac{1}{2} v_0 Q^2_\alpha 
    & 
    \alpha = \beta, 
  \end{array}
\right. 
\label{SIeq:monE}
\eea
This gives the same total energy as Eq.~\ref{SIeq:chargeE} up to an 
unimportant overall constant $(1/2) \sum q^2_i = N (\mu / a_d)^2$. 
Thus, it is also equivalent to the dipolar energy 
Eq.~\ref{SIeq: spin ice energy} up to corrections which are 
small everywhere, and vanish with distance at least as fast as $1/r^{5}_{\ }$
for each dipole pair. 
In particular, one can show that $v_0$ is large enough, for the compounds 
we consider, that it enforces $Q_\alpha = 0$ everywhere in the ground state. 
These are just the ice rules, and the degeneracy of all the states follows 
from the fact that all terms with $r_{\alpha\beta} > 0$ vanish for 
$Q_\alpha \equiv 0$. 
%
%

\end{document}